\documentclass[doublespacing]{elsart}
\usepackage{natbib}
 \usepackage{graphicx}
\usepackage{epstopdf}

\begin{document}
\runauthor{Coward and Regimbau}
\begin{frontmatter}
\title{Detection regimes of the cosmological gravitational wave background from astrophysical sources}
\author[Paestum]{David Coward and}
\author[Rome]{Tania Regimbau}

\address[Paestum]{School of Physics, University of Western
Australia, M013, Crawley WA 6009, Australia, Email: coward@physics.uwa.edu.au}
\address[Rome]{Departement Artemis, UMR 6162
Observatoire de la C\^ote d'Azur, BP 4229
06304 Nice Cedex 4 (France)}

\begin{abstract}
Key targets for gravitational wave (GW) observatories, such as LIGO and the next generation interferometric detector, Advanced LIGO, include core-collapse of massive stars and the final stage of coalescence of compact stellar remnants.  The combined GW signal from such events occurring throughout the Universe will produce an astrophysical GW background (AGB), one that is fundamentally different from the GW background by very early Universe processes.  One can classify contributions to the AGB for different classes of sources based on the strength of the GW emissions from the individual sources, their peak emission frequency, emission duration and their event rate density distribution. This article provides an overview of the detectability regimes of the AGB in the context of current and planned gravitational wave observatories. We show that there are two important AGB signal detection regimes, which we define as `continuous' and `popcorn noise'. We describe how the `popcorn noise'  AGB regime evolves with observation time and we discuss how this feature distinguishes it from the GW background produced from very early Universe processes.
\end{abstract}

\end{frontmatter}

\section{Introduction}
Astronomy has at its disposal a powerful set of technological
tools to explore the cosmos, enabling new windows to the Universe
to be opened. During the 19th and early part of the 20th century
no one envisaged the scope of astrophysical phenomena that the
electromagnetic (EM) spectrum would reveal.  The optical window
into the Universe has been extended to include radio, infrared,
ultraviolet, x-ray and $\gamma$-ray components of the EM spectrum,
and neutrino detectors have been constructed and brought into
successful operation. These technology-based advances have led to
many dramatic discoveries, including the cosmic microwave
background radiation and exotic objects such as neutron-stars and
super-massive black-holes. Gravitational wave (GW) astronomy, a new
window, offers the possibility of dramatically extending our
present understanding of the Universe. This new spectrum is
presently totally unexplored.

Three sets of long baseline laser-interferometer GW detectors have been,
or are nearly, constructed. The US LIGO (Laser Interferometer
Gravitational-wave Observatory) has started its 4th science run (S4), using two
4-km arm detectors situated at Hanford, Washington, and
Livingston, Louisiana; the Hanford detector also contains a 2-km
interferometer. The Italian/French VIRGO project is commissioning
a 3-km baseline instrument at Cascina, near Pisa. There are
detectors being developed at Hannover (the German/British GEO-600
project with a 600-m baseline, commissioning completed May 2006) and near Perth (the Australian International Gravitational
Observatory, AIGO, initially with an 80-m baseline). A detector at
Tokyo (TAMA-300, 300-m baseline) has been in operation since 2001. The
astrophysical detection rates are expected to be low for the
current interferometers, such as `Initial LIGO', but
second-generation observatories with high optical power are in the
early stages of development; these `Advanced' interferometers have
target sensitivities that are predicted to provide a practical
detection rate. 

The currently operating detectors are searching for both astrophysical
sources of GWs, such as merging binary neutron-star and black-hole
systems and core-collapse supernovae, and stochastic backgrounds
composed of many individually unresolved astrophysical sources at
cosmological distances \citep{fer99,cow01,cow02}. 
In addition to these sources, detectors may also be sensitive to a stochastic background of gravitational radiation from the early Universe, a quest that is considered by many to be the `holy grail' of GW searches. 

We provide here a brief sketch of several key ideas underpinning models for the primordial background, but our main focus is on the other background--a stochastic background produced by astrophysical sources distributed throughout the Universe.  This GW signal, as well as being of profound fundamental interest, provides an interesting comparison to that expected from the primordial Universe. 

 \section[Primordial GW background]{Cosmological background of
primordial GWs} Detection of a GW background from the early Universe would have a profound impact on early-Universe cosmology and on high-energy physics, opening up a new window to explore both the Planck epoch at $10^{-43}$ seconds into the
evolution of the Universe and physics at correspondingly high energies that will never be accessible by any other means. Relic
GWs must carry unique information from the primordial plasma at a graviton decoupling time when the Universe had a
temperature $\sim 10^9$ K, providing a `snapshot' of the state of the Universe at that epoch: cosmological GWs
could probe to great depth in the very early Universe.

As currently understood, mechanisms for generating GWs in the primordial Universe can best be described as speculative. Four popular
mechanisms are mentioned briefly here; see Maggiore (2000, Sects. 8, 9 \& 10) for a comprehensive account.

\noindent{\bf Vacuum fluctuations.} In inflationary models, as the Universe cooled, it passed through a phase in
which its energy density was dominated by vacuum energy, and the scale factor increased exponentially fast. The spectrum of GWs that might be observed today from this period results from adiabatically-amplified zero-point energy fluctuations. The spectrum is expected to be extremely broad, covering $10^{-18}-10^8$ Hz.

\noindent{\bf Phase transitions.} Phase transitions in the early Universe, particularly at the electro-weak
scale, are potential sources of GWs. The Standard Model of particle physics predicts that there will be a smooth
cross-over between states rather than a phase transition, but supersymmetric theories predict first-order phase
transitions, hence allowing for possible strong GW emission in the $10^{-5}-1$ Hz band.

\noindent{\bf Cosmic strings.} Grand unified theories predict topological defects; among these, vibrating cosmic
strings, characterized by a mass per unit length, could potentially be strong sources of GWs. The spectrum has a narrow band feature peaking at $10^{-12}$ Hz and a broad band component in the $10^{-10}-10^{10}$ Hz band.

\noindent{\bf Coherent excitations of the Universe}, regarded as a `brane' or surface defect in a
higher-dimensional universe. The excitations could be of a `radion' field that controls the size or curvature of
the additional dimensions, or they could be of the location and shape of our Universe's brane in the higher
dimensions. If there was an equipartition of energy  between these excitations and other forms of energy in the
very early Universe, then the excitations could produce GWs strong enough for detection by advanced LIGOs. They
could probe one or two additional dimensions of size $\sim 10^{-13}$ -- $10^{-16}$ m. If
the number of extra dimensions is more than 2, much smaller scales could be reached (Cutler \& Thorne 2002;
Hogan 2000).
 
 \subsection{Characterizing the background}\label{chartheback}
A cosmological stochastic background of GWs is expected to be isotropic, stationary and unpolarized. The intensity of a stochastic background of GWs is conventionally
characterized by the dimensionless `closure density', $\Omega _{GW}(f)$, defined as the fraction of energy density of GWs
per logarithmic frequency interval normalized to the cosmological critical energy density $\rho_c c^2$ required
to close the Universe:
\begin{equation}
\Omega_{GW}(f)=\frac{1}{\rho_c c^2}\frac{{\rm d}\rho_{\rm GW}}{{\rm d}\log f},
\end{equation}
where $\rho_{\rm GW}(f)$ is the energy density of the background at frequency $f$. In terms of the present value of
the Hubble constant $H_0$, written as $h_0\times100$ km s$^{-1}$ Mpc$^{-1}$:
\begin{equation}
\rho_c  =\frac{3H_0^2}{8\pi G}\approx 1.9 \times 10^{-26} h_0^2 \hspace{0.25cm} \mathrm{kg\hspace{0.15cm}
m^{-3}}.
\end{equation}
\indent There are several other ways of describing the spectrum, including the spectral density (density in
frequency space) of the ensemble average of the Fourier component of the metric\footnote{see Maggiore 2000, Sect 2.2 for a complete derivation of $S_h(f)$}, $S_h(f)$, with units of
Hz$^{-1}$; this allows the experimentalist to express the sensitivity of the detector in terms of a strain
sensitivity, $S_h^{1/2}(f)$, with dimension Hz$^{-1/2}$. Also, the
spectrum can be expressed as a dimensionless characteristic amplitude\footnote{note that the Fourier convention used here is $\tilde{g}(f)=\int_{-\infty}^{\infty} {\rm d}t \,{\rm exp}(2 \pi i f t) g(t)$ so that
$g(t)=\int_{-\infty}^{\infty} {\rm d}f \,{\rm exp}(-2 \pi i f t) \tilde{g}(f)$}. The relationships between $h_c(f)$, $S_h(f)$ and $h_0^2\Omega_{GW}(f)$, as derived by e.g. Maggiore (2000, Sect.
2.2), are:
\begin{equation}\label{1hc}
h_c(f)= \sqrt{2fS_h(f)},
\end{equation}
\begin{eqnarray}\label{1omega1}
h_0^2\Omega_{GW}(f)=(2\pi^2/3) f^2h_c^2(f)=(4\pi^2/3) f^3S_h(f).
\end{eqnarray}
 
 \subsection[Observational bounds]{Observational bounds on $\Omega_{GW}$}
The integrated closure density of GWs from all frequency bands is expressed in dimensionless form as
\begin{equation}
\Omega_{GW} \equiv \int{\Omega_{GW}(f) d(\ln f)}.
\end{equation}
The current best limit for $h_0^2\Omega_{GW}$ is $6\times 10^{-6}$ (Cutler \& Thorne 2002); a value larger than
this would have meant the Universe has expanded too rapidly through the era of primordial nucleosynthesis
(Universe age $\sim $ a few minutes), distorting the universal abundances of light elements away from their
observed values.

Pulsars are natural GW detectors. As a GW passes between us and the pulsar, the time of arrival of the pulse
will fluctuate. The bound on $\Omega_{GW}(f)$ from pulsar timing is (Maggiore 2000, Sect. 7.3):
\begin{equation}
 h_0^2 \Omega_{GW}(f) < 10^{-8},\hspace{0.4cm} f \sim 10^{-8}\hspace{0.1cm} \mathrm{Hz}\hspace{0.1cm}.\\
\end{equation}

A background of GWs will also cause fluctuations in the temperature of the cosmic microwave background. The
bound from anisotropy measurements (Maggiore 2000, Sect. 7.2) is:
\begin{equation}
h_0^2 \Omega_{GW}(f) < 7\times 10^{-11}(H_0/f)^2,\hspace{0.5cm} 1<f/H_0<30, \\
\end{equation}

The proposed Laser Interferometer Space Antenna (LISA) will consist of three spacecraft flying 5 million km apart in
the form of an equilateral triangle. The centre of the triangular formation will be in the ecliptic plane 1 AU
from the Sun and 20 degrees behind the Earth. The main advantage of a space-based interferometer is the removal
of the limitations imposed by gravity gradient and seismic noise, allowing a space-based detector to probe a
low-frequency range that is not accessible to terrestrial detectors. Observations in the very low-frequency
domain 10$^{-4}$ Hz to 0.1 Hz are planned. Because $\Omega_{\rm GW}(f)\propto f^3 S_h(f)$, as in Eq.
(\ref{1omega1}), an order of magnitude decrease in frequency results in a three order of magnitude increase in
sensitivity to $\Omega_{\rm GW}(f)$.

Although this increase in sensitivity is dramatic in the context of cosmological GW backgrounds, it introduces
the problem of sensitivity to the Galactic population of white dwarf binaries. The signal from this Galactic source
may prove to be a limiting factor in searches for cosmological backgrounds.

\section{Astrophysical GW backgrounds (AGB)}\label{AGBs}
 \subsection{Conceptual overview of the AGB}
The emission of GWs from astrophysical sources throughout the Universe can create a stochastic background of
GWs. It could potentially be used to probe the Universe at redshifts $z\sim$ 1--10 and be used as a tool to
study the evolving star formation rate, supernova rates and the mass distribution of black-hole births. However, from the point of view
of detecting the cosmological background produced in the primordial Universe, the astrophysical background is a `noise', which
could possibly mask the relic cosmological signal. Hence, an understanding of AGBs is important on two fronts:
first to provide fundamental knowledge of astrophysical source evolution on a cosmological scale and second to
differentiate this background from the early-Universe background.
 
The astrophysical sources that contribute to a GW background can be broadly categorized as being either
continuous or burst sources. In the context of observing the AGB, continuous GW emission is a somewhat ambiguous
expression, as it depends on the observation time. We provide more rigorous definitions:

{\bf AGB continuous sources}---Astrophysical sources yielding slowly evolving emission with characteristic
evolution time $\tau$ that is very long compared with the observation time $T$.

By choosing $T\approx 1$ yr as a practical observation time, sources that could be considered continuous include
compact binary systems in the gradual in-spiral phase prior to the final few minutes of their evolution and
deformed neutron-stars. Galactic white dwarf binaries are expected to make a significant contribution to the AGB.

{\bf AGB burst sources}---Astrophysical sources with a GW emission time $\tau$ that is
very short compared with the observation time $T$.

\noindent A typical burst source example is a core-collapse supernova, for which $\tau$ could be of order $10^{-3}$ s to minutes.
Others include the late in-spiral and coalescence stages of binary systems, neutron-star bar-mode instabilities and
gamma-ray burst sources.

The total event rate observed in our frame, $R_\mathrm{total}$, is obtained by integrating events in shells of $z$  to the
epoch when the events of the given type first started. For burst sources of a given
type, the fraction of $T$ that contains information on these events is of order $R_\mathrm{total}\times \tau$. Clearly, this
indicates that $R_\mathrm{total}$ introduces an additional constraint for a stochastic background composed of burst sources;
that is $R_\mathrm{total} \tau \approx 1$. The dimensionless quantity $R_\mathrm{total}\tau$ is termed duty cycle ($DC$). A $DC=1$ implies a continuous signal and for a $DC=0.5$, the signal is on average present only $50\%$ of the time . The formal definition of $DC$ that incorporates cosmological sources is:
\begin{equation}\label{DC-ch1}
DC = \int\limits_{0}^{z_c}(1+z) \tau (dR/dz) dz,
\end{equation}
where $\tau$ is time dilated to $(1+z)\tau$ by cosmological expansion and $dR/dz$, the variation of the event
rate with $z$, is dependent on the cosmological model; the integration limit $z_\mathrm{c}$ corresponds to the
epoch when the events of the given type first started.

\subsection{Two AGB Components}
The AGB comprises two components:

\noindent{\bf AGB continuous component}---The contribution to the AGB from sources yielding a $DC\gg 1$, either because
$\tau$ is large or because $R_\mathrm{total}$ is large. The amplitude distribution will be Gaussian as shown by the central limit theorem, which states that the sum of random events from different distributions converges to a Gaussian distribution. Such backgrounds could include the gradual in-spiral phase of compact
binary systems, e.g white-dwarf, neutron-star and black-hole binaries and other systems for which $R_\mathrm{total}$ is large and $\tau$ is not extremely small.

\noindent{\bf AGB `popcorn noise' component}---The contribution to the AGB from burst sources for which
$\tau$ is short but $R_\mathrm{total}$ is high enough that $DC\sim 1$. It will manifest in GW data as a stochastic background comprising mostly non-overlapping individual GW emissions with an amplitude distribution dominated by the spatial distribution of the sources. Possible sources include core-collapse supernovae, neutron-star bar-mode
instabilities and GW emission associated with gamma-ray bursts. This component is characterized by a very skewed 
distribution in GW amplitude. The distribution is found to possess a maximum corresponding to cosmological
sources at a redshift of 2--3 and a long `tail', representing more `local', events.

Different populations of source types (continuous and burst) will superpose to form a composite AGB that may be
very complex. Nonetheless there could be single-source components that dominate at different frequencies, based on the simple fact that the astrophysical source types vary considerably in emission
mechanisms.

\subsection{Spectral properties of the AGB}
The fluence  of a single source located at redshift $z$ is given by
 \begin{equation}
f_{\nu _o}=\frac{1}{4\pi d_L^2}\frac{dE_{GW}}{d\nu }(1+z)
\label{eq-fluence}
\end{equation}
where $d_{L}=(1+z)r$ is the luminosity distance, $r$
is the proper distance, 
${dE_{GW}}/{d\nu}$ is the angular averaged GW energy spectrum,
$\nu=(1+z)\nu _{o}$ the frequency in the source frame and
$dR(z)$ is the event rate at redshift $z$. 

The integrated flux at the observed frequency $\nu_o$ is defined by:
\begin{equation}
F_{\nu _{o}}=\int f_{\nu _o}dR(z)
\end{equation} 

Using the definitions above,  a stochastic gravitational wave background of astrophysical origin, eq.(1) can be expressed as
\begin{equation}
\Omega _{GW}(\nu _o)=\frac{1}{c^3 \rho _c}{\nu _o}F_{\nu _o}
\end{equation}
  
\begin{figure}
\includegraphics[scale=0.8]{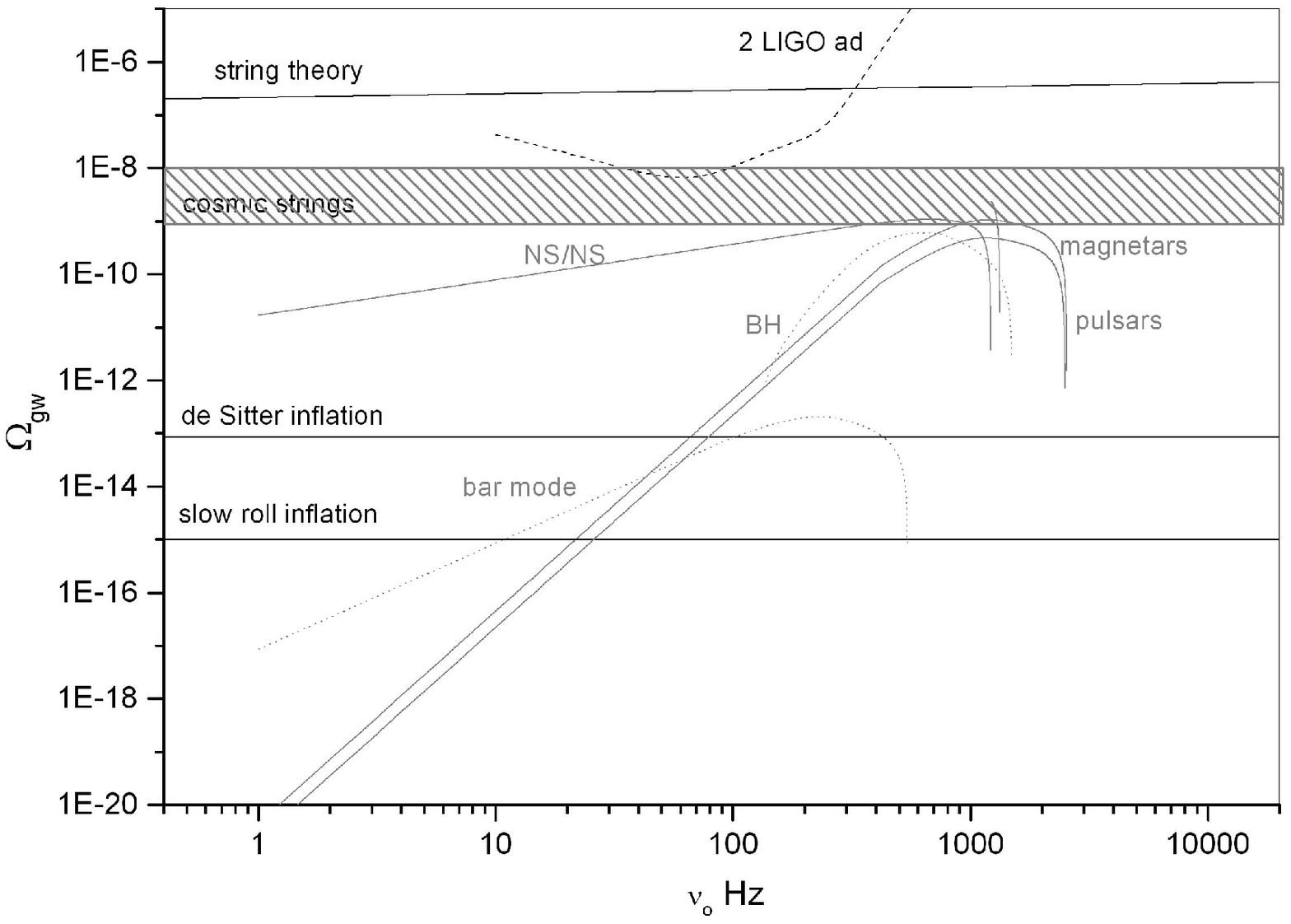} \caption{ The dimensionless gravitational wave closure density, $\Omega_{gw}$, plotted as a function of observed frequency for potentially significant contributions from astrophysical GW sources: continuous background (continuous line) such as from neutron-star--neutron-star coalescence (Regimbau \&
de Freitas Pacheco 2005b), rotating
pulsars or magnetars (Regimbau \&
de Freitas Pacheco 2005a) and shot
noise (dotted line) such as black-hole ringdown and bar-mode instabilities in young, fast rotating neutron-stars (Regimbau 2005b).   
Theoretical limits for several primordial GW background
models are plotted (horizontal lines) and the detection sensitivity from correlating 
two Advanced LIGO type interferometers is shown (dashed line).}\label{agb}
\end{figure}

Different AGB models have been investigated (see Regimbau
2005 for a complete review) by using several astrophysical sources of GWs.
On the one hand, distorted black-holes (Ferrari et al. 1999; de Araujo et al. 2000) are examples of sources able to generate a shot
noise (time interval between events large in comparison with duration of a single event), while supernovae or hypernovae (Blair et al. 1996;
Coward et al. 2001; Coward et al. 2002 ; Howell et al. 2004; Buonanno et al. 2004) are expected to produce an intermediate `popcorn'
noise i.e. between a shot noise and continuous signal. On the other hand, the contribution of tri-axial rotating neutron-stars (Regimbau \& de Freitas Pacheco 2001), 
including magnetars (Regimbau \& de Freitas Pacheco 2005a), constitutes
a truly continuous background.
Figure (\ref{agb}), from Reigimbau 2005, shows different types of AGB spectra along with spectra of a primordial cosmological background.

\subsection{Continuous AGB to single event}
The AGBs composed of burst sources discussed above are assumed to
consist of a continuous (or nearly continuous) background of GWs
that would be orders of magnitude below the noise level of any
single Advanced LIGO-type detector.  Qualitatively one expects
that the AGB will be dominated by sources typically 3--4 orders of
magnitude fainter in luminosity than sources in relatively nearby
galaxies at 10--100 Mpc ($z\approx$ 0.002--0.02).

There is a crucial difference between a primordial cosmological GW background and the AGB: the primordial
background is long-term stationary over any practical observation time $T$ but, in contrast, the AGB signal over
$T$ will fluctuate according to $dR/dz$. To explain the latter point, consider GW events associated with stellar
core collapses. Assume these events occur at a local rate of 1 yr$^{-1}$ out to 15 Mpc and produce a signal above
the detectable threshold in an idealized GW detector at this source distance. For $T\ll 1$ yr the detector will be
dominated by a roughly stationary background from sources at $z\approx$ 1--2, well below the detectable
threshold. But, as $T\rightarrow 1$ yr, it will be highly probable that the GW detector will record a local
event. What is observed as a local event can be interpreted as the low probability tail of $dR/dz$ or the local
component of the AGB.

The low probability tail contains likely sources that may be
detectable as single events by LIGO and Advanced LIGO
over observation times of about 1 yr:\\

\emph{A significant part of current detection strategies are focussed on trying to detect the low probability
tail of the popcorn noise distribution.}\\

The high-$z$ sources will accumulate as unresolved signals in the
data and with no prior knowledge of this background, it will manifest as an excess noise in the detector. The optimal signal detection method is cross-correlating two or more detectors.

\section{The probability event horizon}
For any
GW detector and GW source type, one can define a
`detectability horizon' centred on the detector and surrounding
the volume in which such events are potentially detectable; the
horizon distance is determined by the flux limit of the detector
and the source flux. A second horizon, defined by the minimum
distance for at least one event to occur over some observation
time, with probability above some selected threshold, can also be
defined. We call this the probability event horizon, or the PEH\footnote{see Coward \& Burman (2005) for a more detailed description of the PEH in a broader astronomical context. The authors apply it to gamma-ray burst red-shift data to demonstrate how it can be used to constrain the local rate density of GRBs.}. It
describes how an observer and all potentially detectable
events of a particular type are related via a
probability event distribution encompassing all such events. The PEH is a way of quantitatively describing how the popcorn noise component of the AGB will manifest in GW detector data.

The idea can be illustrated using a static Euclidean universe, assuming an
isotropic and homogenous distribution of events. For a constant
event rate $r_0$ per unit volume, the mean cumulative event rate
in a volume of radius $r$ is $ R(r)=(4/3)\pi r^{3}r_0$. The events
are independent of each other, so their distribution is a Poisson
process in time: the probability for at least one event to occur within the radius $r$ is:
\begin{equation}\label{prob1}
p(n\ge1;R(r),T) = 1 - e^{-R(r) T}= \epsilon\,,
\end{equation}
$e^{-RT}$ being the probability of zero events occurring. 
For a chosen value of $p$, one solves the equation $p=\epsilon$ for $r$, the
corresponding radial distance. The PEH is defined as:
\begin{equation}\label{horizon}
r_{\epsilon}^{\mathrm{PEH}}(T)= (3N_{\epsilon}/4\pi r_0)^{1/3}
T^{-1/3}\;.
\end{equation}
The speed at which the horizon approaches the observer---the PEH
velocity---is obtained by differentiating $r_{\epsilon}(T)$ with
respect to $T$:
\begin{equation}\label{velocity}
v_{\epsilon}^{\mathrm{PEH}}(T)=(N_{\epsilon}/36 \pi r_0)
^{1/3}T^{-4/3}\,.
\end{equation}

\begin{figure}
\includegraphics[scale=0.8]{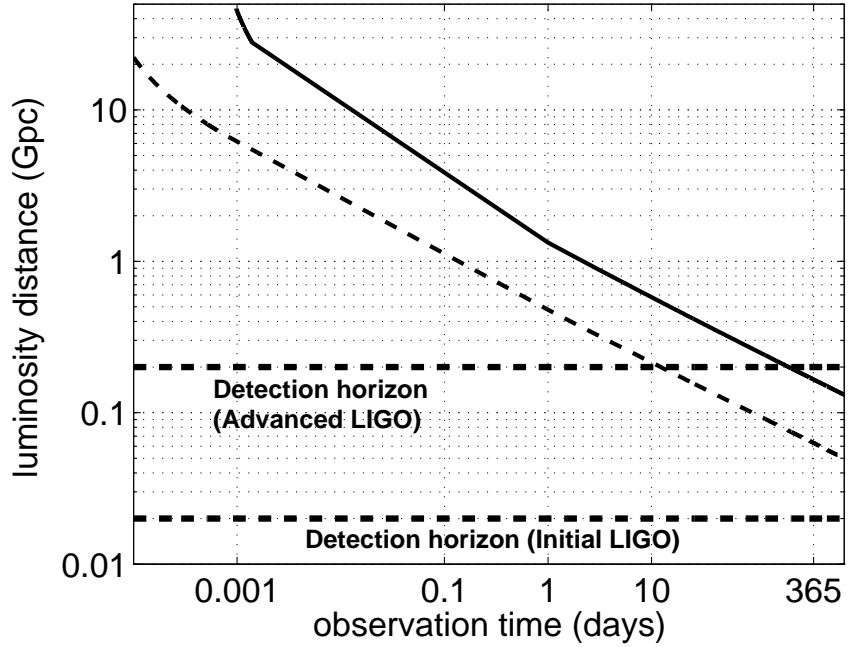} \caption{The PEH evolution for double neutron-star mergers, from Coward et al. (2005), as a function of observation time.  Galactic merger
rate limits are very uncertain, but we use  292 Myr$^{-1}$ (dashed
curve) and 17 Myr$^{-1}$ (solid line) as examples, and we assume the merger rate evolution
following the evolving SFR. The horizontal lines show the sensitivity limits for Advanced and Initial LIGO and the PEH curves assume an idealized optimal detection filter. }\label{peh1}
\end{figure}

The PEH concept is a useful tool for modelling the popcorn component of the AGB, such as the background from double neutron-star (DNS) mergers. A PEH model can in principle be used to constrain the local rate density of events in GW detector data, as shown by Howell et al. (2006). It  also provides a way of describing the detectability of the AGB for a certain GW detector sensitivity, such as LIGO or Advanced LIGO (Coward et al. 2005). Figure \ref{peh1} plots the PEH for DNS mergers, assuming the events are roughly proportional to the evolving star-formation rate (SFR). Coward et al. (2005) justify this assumption by showing that a 1--2 Gyr merger time does not significantly
alter the distribution of mergers in redshift, especially for
small $z$.  The plot shows how the horizon approaches the detectability volume encompassing the LIGO and Advanced LIGO detectors. The time evolution of the AGB described by the PEH shows how the rate density of events goes from a high rate of distant low-luminosity events to rare higher luminosity events.

\section{Detection regimes of the AGB}
If the number of GW sources is
large enough that the time interval between events, as observed in our frame, is small compared
to the duration of a single event $(DC >> 1)$, the combined GW emissions will
produce a continuous background. The central limit theorem predicts that this regime of the
AGB  will yield a Gaussian signal. For this case the optimal detection 
strategy is to cross-correlate the output of two (or more) detectors, assuming they have 
independent spectral noises. Matched filtering may also be used here to optimise the cross-correlation.
 
In the AGB regime where the source rate density is small and the
time interval between events is long compared to the duration of a
single event $(DC << 1)$, the sources are resolved and the nearest may be
detected by `burst' data analysis techniques. The optimal method is
matched filtering but this filter can only be optimal if the exact shape of
the signal is known. Robust generic methods based on coincident analysis are
also being developed to search for events where the exact form of the signal is not known (see for instance
Arnaud et al. 1999; Pradier et al. 2001; Sylvestre 2002; Chassande Mottin \& Pai 2005).

In the intermediate `popcorn noise' regime, where the time interval
between events is of the same order as the duration of a single event,
the waveforms may overlap but the amplitude distribution will not be Gaussian. The amplitude distribution of this background signal will contain information on the spatial distribution of the sources. This point is highlighted by using the PEH description for `popcorn noise' :  the amplitude of the individual events scales with distance and the distance to the largest-amplitude event is reducing as a function of observation time, as shown in figure \ref{peh1}. The time dependence of the signal is important and data analysis techniques in the frequency domain, such as the cross correlation
statistic, do not utilze this feature of the signal.
New data analysis techniques are currently under
investigation, such as the search for anisotropies (Allen and Ottewill, 1997) which can be used to create a map of the GW background (Cornish 2001; Ballmer 2005), the maximum
likelihood statistic (Drasco \& Flannagan 2003), 
or by utilizing the PEH concept described
in the previous section (Howell et al. 2006).

The different regimes of the AGB are being studied using population synthesis. Regimbau \& de Freitas Pacheco (2005b) performed
numerical simulations of the cosmological population of coalescing
neutron-star--neutron-star binaries. Figure 3 shows that events located beyond the critical redshift $z_*
= 0.23$ produced a continuous background, while those in the redshift
interval $0.027<z<0.23$ produce a nearly continuous (popcorn noise) signal. The signal is discrete, i.e. shot noise like, for events occurring closer than $z = 0.027$. 
The GW closure density, $\Omega_{gw}$, for the continuous component, has a maximum
at 670 Hz with an amplitude of $1.1\times 10^{-9}$, while the popcorn
noise component has an amplitude about one order of magnitude higher with maximum at 1.2 kHz. That study highlights the importance of the popcorn noise regime to the AGB. 
\section{Discussion}
We show that the AGB can be classified into several detection regimes. For continuous GW backgrounds, cross correlating multiple detectors is the optimal search method, assuming that the noise in each detector is not correlated. Examples of GW source populations where this technique is appropriate include both Galactic binaries and massive cosmological binary systems. The central limit theorem predicts that the sum of the GW signal from these sources produces an amplitude distribution that is Gaussian. Because of the inverse distance dependence for the GW amplitude, this continuous regime may be dominated by a small number of relatively close systems, implying that matched filtering will be the optimal search method. This is an example of how the low probability tail of the GW event rate density, corresponding to nearby sources, dominates the signal over `long' observation times.

We identify another AGB regime, where the GW signal composed of `burst souces' contains information on the source-rate distribution. An example of the sources contributing to such a background include the GW emission associated with gamma-ray bursts and massive stellar collapse. Even if the cosmological event rate for these sources is high enough to produce a truly continuous GW background, there is a horizon distance where the signal will be dominated by nearby events. Hence it is apparent that there is a transition from a small amplitude continuous unresolved signal to higher-amplitude discrete events. 

The detectability of the AGB is determined by the detector sensitivity which sets the detection horizon. If the detector horizon extends to distances corresponding to event rates for which the received signal is nearly continuous, i.e. to the popcorn noise regime, it is possible to probe the source distribution and its association with the evolving star formation rate. Because the present detector horizon for LIGO is of some tens to hundreds of Mpc for the strongest proposed GW sources, the AGB regime currently probed is dominated by rare single events corresponding to relatively long observation times, as shown by the PEH curves in Figure 2. As the sensitivity of LIGO improves and the next generation of GW observatories come on line,  it is plausible that the AGB can be explored for the first time, providing new insight into both the sources generating the AGB and the evolutionary history of the Universe.

\section{Acknowledgements}
This article evolved from discussions during 2005 conducted at the Observatoire de la C\^ote d'Azur, Nice, and a workshop in Perth, Australia, sponsored by the Australian Consortium for Interferometric Gravitational Astronomy (ACIGA).  T. Regimbau thanks ACIGA for providing support during this workshop. D. Coward thanks Dr R. Burman and Prof. D. Blair for providing feedback during development of the `probability event horizon' model. The authors thank the referee for providing invaluable advice and corrections to the manuscript. D. M. Coward is supported by an Australian Research Council Research Fellowship.

\newpage

\begin{figure}
\includegraphics[scale=0.8]{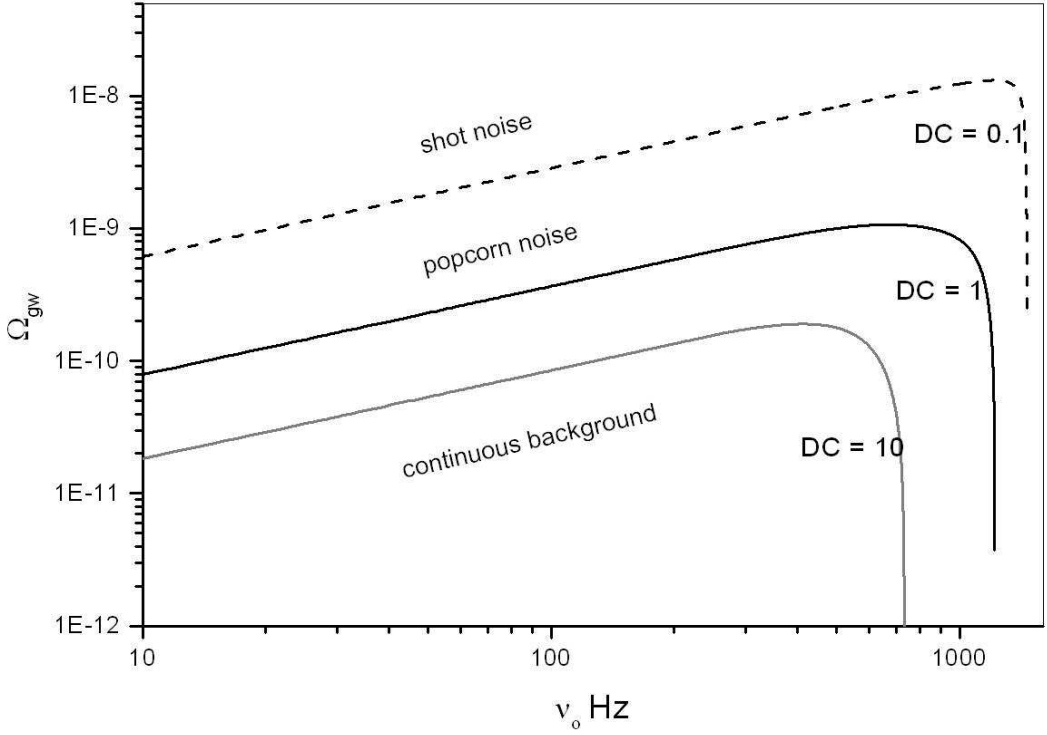} \caption{Same as Figure 1, but showing the different regimes of the AGB for double neutron-star mergers
  occurring beyond $z_* = 0.23$ (bold continuous curve). The 
regime with duty cycle $DC>10$, corresponds to sources beyond
  $z=1.05$ (grey continuous curve). The
  `popcorn' noise regime, with $DC=1$, corresponds to sources in the redshift interval $0.027<z<0.23$ (solid line), and the `shot' noise regime, with $DC<0.1$, corresponding to sources at $z<0.027$. The GW closure density is dominated by the popcorn and shot noise contributions. }\label{omega}
\end{figure}

\end{document}